\documentclass[fleqn,usenatbib]{mnras}
\usepackage{newtxtext,newtxmath}
\usepackage[T1]{fontenc}
\usepackage{ae,aecompl}
\usepackage{rotating}
\usepackage{graphicx}	
\usepackage{amsmath}	
\usepackage{amssymb}	

\title[Experiments on cometary activity: ejection of dust aggregates from a sublimating water-ice surface]{Experiments on cometary activity: ejection of dust aggregates from a sublimating water-ice surface}

\author[D. Bischoff et al.]{
D. Bischoff,\thanks{E-mail: d.bischoff@tu-bs.de}
B. Gundlach,
M. Neuhaus,
J. Blum\\
Institut f\"ur Geophysik und extraterrestrische Physik, Technische Universit\"at Braunschweig, Mendelssohnstr. 38106 Braunschweig, Germany
}

\date{}

\pubyear{2018}

\begin{document}
\label{firstpage}
\pagerange{\pageref{firstpage}--\pageref{lastpage}}
\maketitle

\begin{abstract}
The gas-driven dust activity of comets is still an unresolved question in cometary science. In the past, it was believed that comets are dirty snowballs and that the dust is ejected when the ice retreats. However, thanks to the various space missions to comets, it has become evident that comets have a much higher dust-to-ice ratio than previously thought and that most of the dust mass is ejected in large particles.
\par
Here we report on new comet-simulation experiments dedicated to the study of the ejection of dust aggregates caused by the sublimation of solid water ice. We find that dust ejection exactly occurs when the pressure of the water vapor above the ice surface exceeds the tensile strength plus the gravitational load of the covering dust layer. Furthermore, we observed the ejection of clusters of dust aggregates, whose sizes increase with increasing thickness of the ice-covering dust-aggregate layer. In addition, the trajectories of the ejected aggregates suggest that most of the aggregates obtained a non-vanishing initial velocity from the ejection event.
\end{abstract}

\begin{keywords}
comets: general - methods: laboratory: solid state
\end{keywords}

\section{Introduction}\label{Sect_1}
\begin{figure}
    \includegraphics[angle=0,width=\columnwidth]{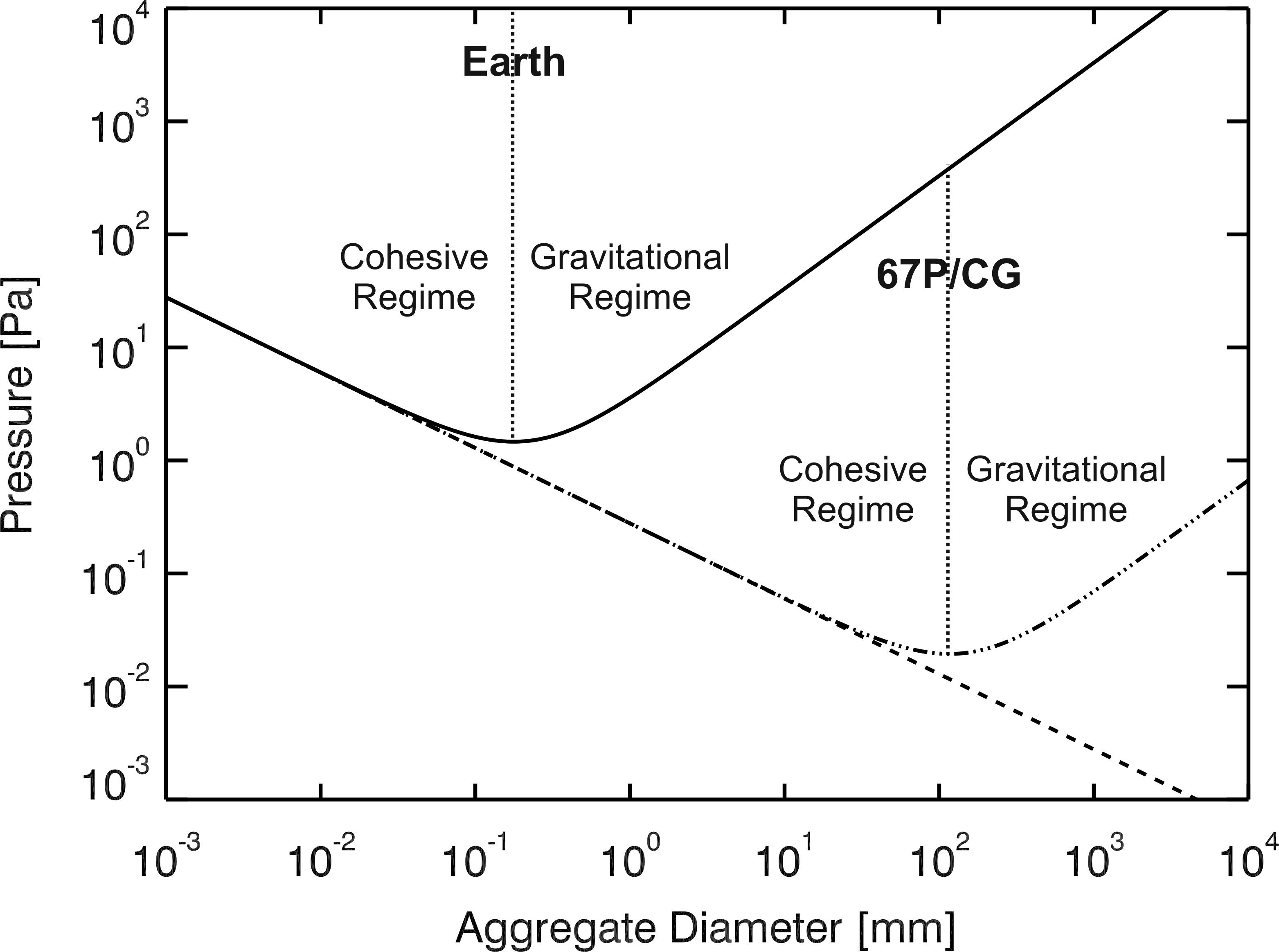}
    \caption{Tensile strength (dashed curve) and gravitational pressure derived for the situation on Earth (solid curve) and on comet 67P/Churyumov-Gerasimenko (dotted-dashed curve, with $g=2\cdot 10^{-4}\,\mathrm{m/}\mathrm{s}^2\,$\citep{Groussin2015}) of a monolayer of dust aggregates as a function of the aggregate radius. Cohesion between the aggregates becomes dominant for aggregates smaller than $\sim 0.1 \, \mathrm{mm}$ in diameter on Earth and for aggregates smaller than several centimeters on comet 67P/Churyumov-Gerasimenko, respectively.}
    \label{fig_1}
\end{figure}
\begin{figure*}
    \includegraphics[width=0.7\textwidth]{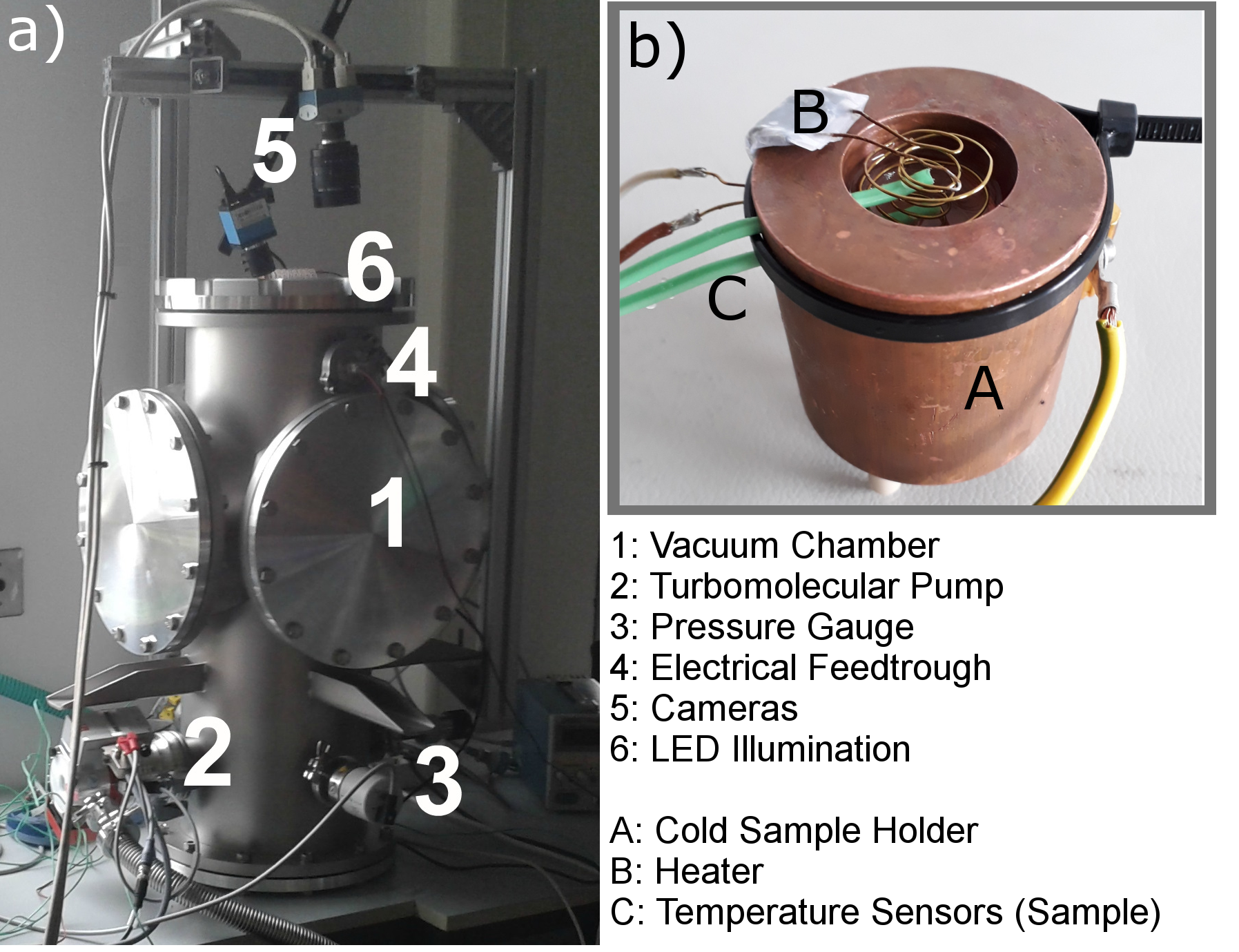}
    \caption{Photographs of the experimental setup (a) and the sample holder (b) that is located inside the vacuum chamber during the experiments. The different parts of the experiment are described in the figure legend.}
    \label{fig_2}
\end{figure*}
Comets are kilometer-sized objects composed of ices (mainly $\mathrm{H_2O}$, $\mathrm{CO_2}$ and CO ice) and dust \citep[dominated by silicates and organic materials,][]{Greenberg1990,Greenberg1998,Levasseur-Regourd2018}. Because of their very low albedo \citep[see, e.g.,][]{AHearn2005a}, comets are among the darkest objects in the Solar System. Hence, the solar radiation can effectively heat the surface of the cometary nucleus, which leads to the sublimation of the volatile constituents and, therewith, to the ejection of surface material.
\par
It has been argued that comets formed in the young Solar System by the gentle gravitational collapse of dust clouds, typically consisting of mm- to cm-sized aggregates \citep{Johansen2007,Skorov2012,Blum2014,Blum_2017}. Due to the nature of this formation scenario and the high dust-to-ice mass ratio of $\sim 4 - 9$ \citep{Lorek2016,Fulle2016}, the comet nucleus is composed mainly of non-volatile dust aggregates, with only minor contributions (in mass) by volatile material.
\par
The nuclei and the surfaces of comets apparently consist of intact dust aggregates \citep{Fulle_Blum2017,Blum_2017} which survived the comet formation process, owing to the small impact velocities during the collapse \citep{Jansson2014,Syed2017,WahlbergJansson2017}. For most of their lifetime since formation, the cometary precursors orbited the Sun at large heliocentric distances so that they remained almost unaffected by solar radiation and mutual collisions \citep{Fulle_Blum2017,Schwartz2018b}. However, gravitational disturbances by the giant planets can change their orbital parameters over time and, thus, bodies from the Kuiper-Belt region can be scattered into the inner Solar System.
\par
Getting closer to the Sun, solar illumination leads to the sublimation of water ice and other volatile species and, thus, to the formation of a volatile-free surface. The desiccated dust-aggregate layers possess a very low thermal conductivity \citep{Gundlach2012}, a low gas permeability \citep{Gundlach2011} and a low tensile strength \citep{Skorov2012,Blum2014,Blum_2017,Attree2018a}. As a result, the sublimation of the volatile constituents underneath the dust layer can lead to the ejection of dust aggregates from the surface \citep{Skorov2012,Gundlach2015}. In fact, most of the ejected dust mass seems to be in bodies exceeding sizes of $\sim 1$ mm \citep{Fulle2016b,Blum_2017,ott2017}, instead of $\mathrm{\mu m}$-sized particles, as might be expected from the existence of the cometary dust tail, which mainly consists of micrometer-sized particles \citep{sekanina1996}. Dust-aggregate release from the surface is possible if the gas pressure underneath the dust cover is sufficient to overcome the sum of the gravitational stress, $p_g$, exerted by the dust-aggregate layer and its tensile strength, $p_t$. The gas pressure can be approximated by the sublimation pressure $p_s$ and by taking into account the influence of the dust layer on the outward diffusion of the molecules. If $\xi$ is the fraction of molecules able to escape through the dust layer into space, $1 - \xi$ is the fraction of molecules scattered back towards the dust-ice interface, which causes a pressure build-up inside the dust layer \citep[see][for a detailed description]{Gundlach2011}. Hence, the condition for dust release can be written as
\begin{equation}\label{Eqation_1}
p_s \, (1 - \xi)\, \geq \,   p_g \, + \, p_t \ \mathrm{.}
\end{equation}
The sublimation pressure reads
\begin{equation}\label{Eqation_1b}
p_s  =  a ~ \mathrm{exp}(-b/T),
\end{equation}
where $T$ is the temperature of the ice. The parameters $a=3.23^{+0.73}_{-0.38}\, 10^{12}\, \mathrm{Pa}$ and $b=6134.6 \pm 17.0\, \mathrm{K} $ are material constants \citep[given for water ice by][]{Gundlach2011}. The gravitational pressure,

\begin{equation}\label{Eqation_2}
p_g = \rho ~ g ~ h,
\end{equation}
depends on the mass density of the dust-aggregate layer $\rho$, on its thickness $h$, and on the local gravitational acceleration $g$, respectively. The mass density of the dust-aggregate layer, in turn, can be derived from the material density $\rho_0$ of the dust grains and from the packing density inside the dust aggregates $\phi_{\rm agg}=0.35$ \citep[see][]{Weidling2012} and inside the dust-aggregate layer $\phi_{\rm pack}=0.6$ \citep[see][]{Skorov2012} by $\rho = \rho_0 ~ \phi_{\rm agg} ~ \phi_{\rm pack}$.
\par
The tensile strength of a layer of dust aggregates of radius $s$ is given by

\begin{equation}\label{Eqation_3}
p_t = p_{t,0} ~ \phi_{\rm pack} ~ s^{-2/3},
\end{equation}
with $p_{t,0} = 1.6\, \mathrm{Pa}$ being an empirical constant when $s$ is given in units of mm \citep{Skorov2012,Blum2014}. A more general treatment of cohesive strength of asteroidal regolith can be found in \citet{Scheeres2010}. However, the foundations of the cohesive strength for both materials, solid regolith particles as well as cometary aggregates, is the van der Waals force between particles in contact (see \citet{Scheeres2010} and \citet{Skorov2012} and references therein).

The larger the aggregates, the smaller the tensile strength and hence the easier to eject the dust from the surface (for low gravity environments such as comets; dashed curve in Fig. \ref{fig_1}). Dust aggregates can be ejected by the sublimation of water ice (or any other volatile material) if the temperature (and, thus, the sublimation pressure) exceeds a critical value, which we will refer to as ``activity temperature'' in this manuscript. In this case, the gas pressure overcomes the tensile strength of the material plus the gravitational pressure of the dust layer. On Earth, the gravitational pressure of one aggregate layer is higher than the tensile strength between the aggregates for $d\, \tiny{\gtrsim} \, 0.2 \, \mathrm{mm}$ (where d is the diameter of the aggregates, see solid curve in Fig. \ref{fig_1}). In the case of comet 67P/Churyumov-Gerasimenko, this condition is fulfilled for dust-aggregate radii of several centimeters, due to the reduced gravitational acceleration of this object. Here, we adopted the value of $g=2\cdot 10^{-4}\,\mathrm{m/}\mathrm{s}^2$ from \citet{Groussin2015} for the surface acceleration of comet 67P.

\par
Unfortunately, the availability of laboratory experiments on the gas-driven dust activity of comets in the literature is scarce. Although several experiments studied the heat and mass transfer inside cometary analogue samples \citep[see, e.g.,][]{Dobrovolskii1977,Storrs1988,Green1999,Bar-Nun2009,Pat-El2009,Koemle1991,Kossacki1999,
Kaufmann2007,Brown2012a,Gundlach2011,Gundlach2012,Blum2014,Poch2016},
the dust ejection caused by the sublimation of volatiles was only investigated in the framework of the so-called KOSI (``KOmeten-SImulation'') experiments \citep{Gruen1991,Gruen1992f,Thiel1989}. \citet{Laemmerzahl1995} observed the ejection of dust particles due to the sublimation of the underlying volatile constituents, followed by the formation of a covering non-volatile dust layer with a low thermal conductivity \citep{Spohn1989d}. This dust layer quickly inhibited the activity of the samples as its thickness reached a certain threshold. Similar observations were made by \citet{Ratke1989} and \citet{Thiel1991}.
\par
The laboratory experiments performed so far were very useful to understand the nature of analogue samples under cometary-like conditions. However, since the Rosetta mission escorted a comet for about two years, the picture of cometary activity has changed. The most important changes relevant for this work are that comets most probably consist of millimeter- to centimeter-sized dust aggregates \citep{Blum_2017} and that comets possess a very high dust-to-ice ratio of $\sim 4-9$, much higher than used in the KOSI experiments \citep{Lorek2016,Fulle2016}. Thus, new comet simulation experiments with realistic comet analogue materials are required to investigate the gas-driven dust activity of comets. First attempts to study the evolution of more realistic ice-dust mixtures under vacuum conditions were made by the planetary science group in Bern \citep[see, e.g.][]{Poch2016}. They studied the influence of organic materials on the sublimation process. One of their main findings is that the mixing type influences the outgassing, the size of the ejected dust aggregates and the appearance of the residuals.
\par
With this work, we intend to establish a new series of comet simulation experiments, concentrating on the investigation of the gas-driven dust activity.
We will focus on the following scientific questions:
\begin{itemize}
\item[1.] Does dust activity occur when the pressure under the dust exceeds the gravitational and cohesive pressure from above as discussed in the literature \citep[see e.g.][]{Kuehrt1994,Skorov2012,Blum2014,Blum_2017,Gundlach2015}?

\item[2.] What is the size of the emitted dust "chunks", i.e., single dust aggregates or clusters thereof?

\item[3.] What is the initial ejection velocity of the dust chunks \citep{Kramer2015}?
\end{itemize}
Our strategy is to start as simple as possible by the investigation of sublimating ice-dust mixtures that can be fully described by established physics. The easiest system imaginable is a solid block of water ice (in hexagonal form) and dust aggregates placed on top of this ice block.
It should be clear that this type of sandwich structure does not represent a real comet, like we also did not use different materials or ices, but it is a simple physical system that can be used to address processes on cometary nuclei. In particular, solid ice means a local dust to ice ratio of zero, which is certainly much smaller than found on comets. However, our dust emission model (Equations \ref{Eqation_1}-\ref{Eqation_3}) uses the gas pressure above the subliming water ice, which is a function of temperature (see Equation \ref{Eqation_1}) and properties of the overlaying dust layers (through the parameter $\xi$ in Equation \ref{Eqation_1b}) only and on first order independent of the dust to ice ratio. It should also be mentioned here,  that real comets are way more complex than our simple model system and that there are other parameters, which may influence dust activity. We will get back to this point in Section \ref{Sect_4}.\\
Thus, we constructed a new experimental setup that will be described in Section \ref{Sect_2} of this paper. We performed a series of experiments with different dust-aggregate sizes and dust-aggregate-layer thicknesses (see Section \ref{Sect_3}). The results of these experiments and their implications for our understanding of cometary activity will be presented in Section \ref{Sect_4}.

\section{Experimental}\label{Sect_2}
In the following we present the experimental setup used to study the dust ejection of sublimating water-ice samples. In addition, the experimental procedure and the calibration measurements are described.
\subsection{Setup}
\begin{figure}
    \includegraphics[width=\columnwidth]{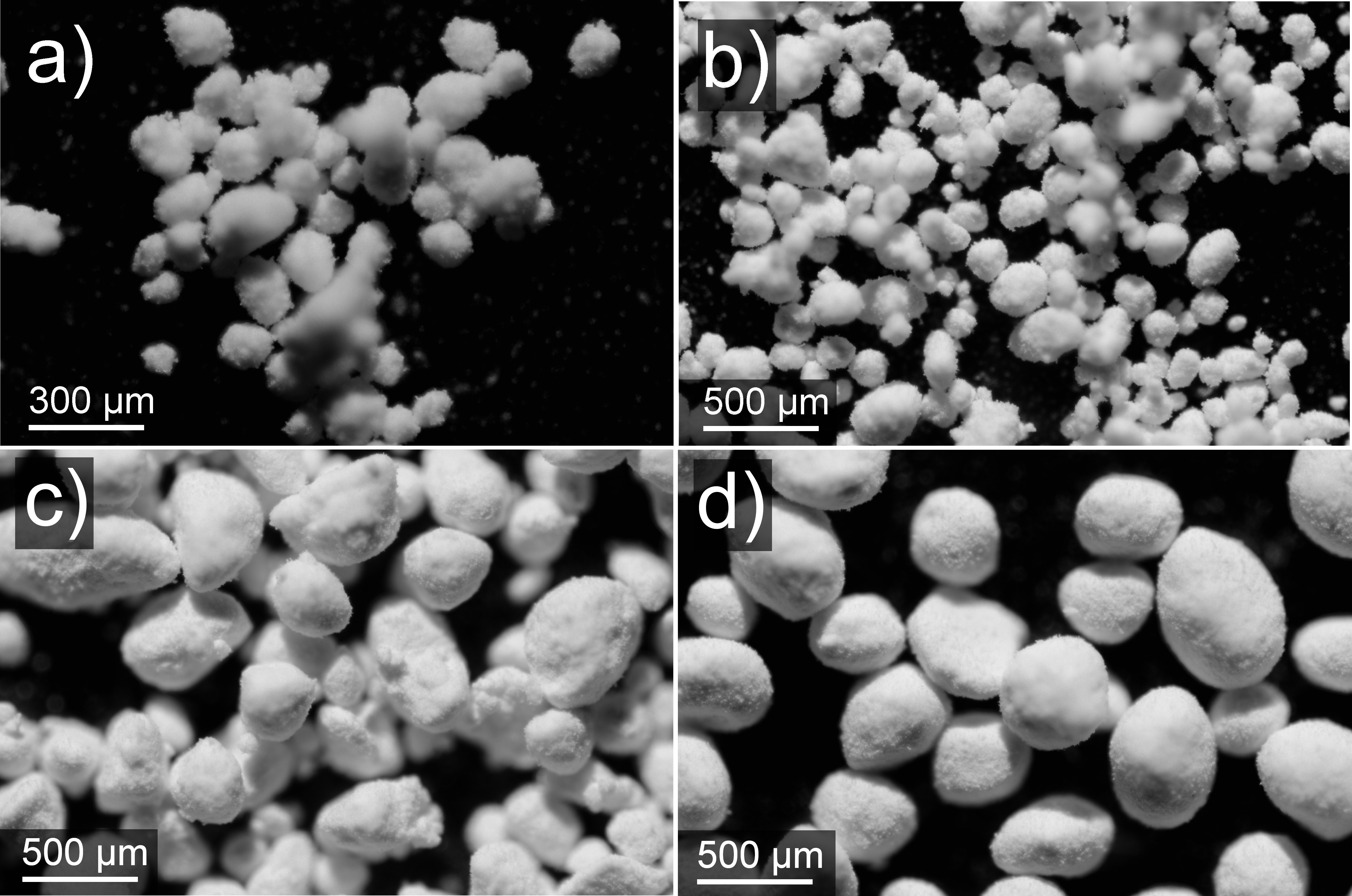}
    \caption{Examples of the silica dust aggregates after sifting into the four size ranges (see Table \ref{size_distribution} for details).}
    \label{fig_3}
\end{figure}

\begin{table}
\centering
\caption{Mesh dimensions, resulting mean dust-aggregate diameters and standard deviations for the used dust-aggregate samples.}
\label{size_distribution}
\begin{tabular}{|c|c|c|c|}
\hline
Figure   & Mesh Size  & Average  & Standard  \\
\ref{fig_3} & {[$\mathrm{mm}$]} & Diameter {[$\mathrm{mm}$]} & Deviation {[$\mathrm{mm}$]} \\
\hline
a) & 0.050-0.100    & 0.119   & 0.031 \\ \hline
b) & 0.100-0.250   & 0.173   & 0.048 \\ \hline
c) & 0.250-0.400   & 0.338   & 0.088 \\ \hline
d) & 0.400-0.500   & 0.440   & 0.079 \\ \hline
\end{tabular}
\end{table}

The experiments were performed inside a vacuum chamber to ensure cometary-like conditions. The ambient gas pressure typically was between $1\, \mathrm{and}\,\, 10\, \mathrm{Pa}$ and the initial temperature of the ice was set to about $190\, \mathrm{K}$ by pre-cooling the sample container. Figure \ref{fig_2}a shows a photograph of the setup. A turbomulecular pump (no. 2 in Figure \ref{fig_2}a) together with a rotary-vane pump were used to provide the low pressure environment, which was monitored by a pressure gauge (no. 3). Two electrical feedthroughs (no. 4) provided power for the heater (B in Figure \ref{fig_2}b) and allowed the measurement of the ice temperature by thermocouple temperature sensors (C). Two cameras (no. 5) were used to monitor the surface evolution of the dust-covered water ice from top and from the side (by using a mirror). The samples were produced (see Section \ref{Sample_Preparation} for details) inside the sample holder (see Figure \ref{fig_2}b), a cylindrical copper block (A), which ensured low temperatures during the experimental runs, when pre-cooled with liquid nitrogen prior to the experiments. A heater (B) was inserted into the ice in order to heat the sample to the required temperatures during the experimental runs.

\subsection{Sample Preparation}\label{Sample_Preparation}
As cometary dust analogues, we chose aggregates consisting of irregular silica ($\rm SiO_2$) monomers with diameters between $\sim 0.1~\rm \mu m$ and $\sim 10~\rm \mu m$ \citep{Kothe2013c} and material density $\rho_0=2.6~\mathrm{g~cm^{-3}}$. The silica aggregates with a spheroidal shape were sieved into different size ranges, between $\sim 0.050$~mm and $\sim 0.500$~mm in diameter. Figure \ref{fig_3} shows examples of and Table \ref{size_distribution} presents details about the four dust-aggregate size ranges.

To prepare a sample, distilled water was poured into the sample holder (see Figure \ref{fig_2}b) so that the heating wire and the temperature sensors were fully covered by the water. Then, the sample was frozen inside a freezer. Due to the expansion of the ice during the freezing process, the final water-ice surface was located slightly above the rim of the sample holder. We then used a warm metal plate to flatten the ice surface. After that, the silica aggregates were carefully sieved onto the ice surface. Due to the sifting process, the forming dust-aggregate layers were quite homogeneous. Different layer heights between one and approximately eight monolayers (in units of the respective dust-aggregate diameters) were achieved by this method. The layer thickness was measured with a photographic image taken from the side of the sample. For this purpose, an edge was carved into the ice and a mirror was placed inside the vacuum chamber so that one of the cameras imaged the sample in a direction parallel to the surface.

\subsection{Experimental Procedure}

\begin{figure}
	\includegraphics[width=\columnwidth]{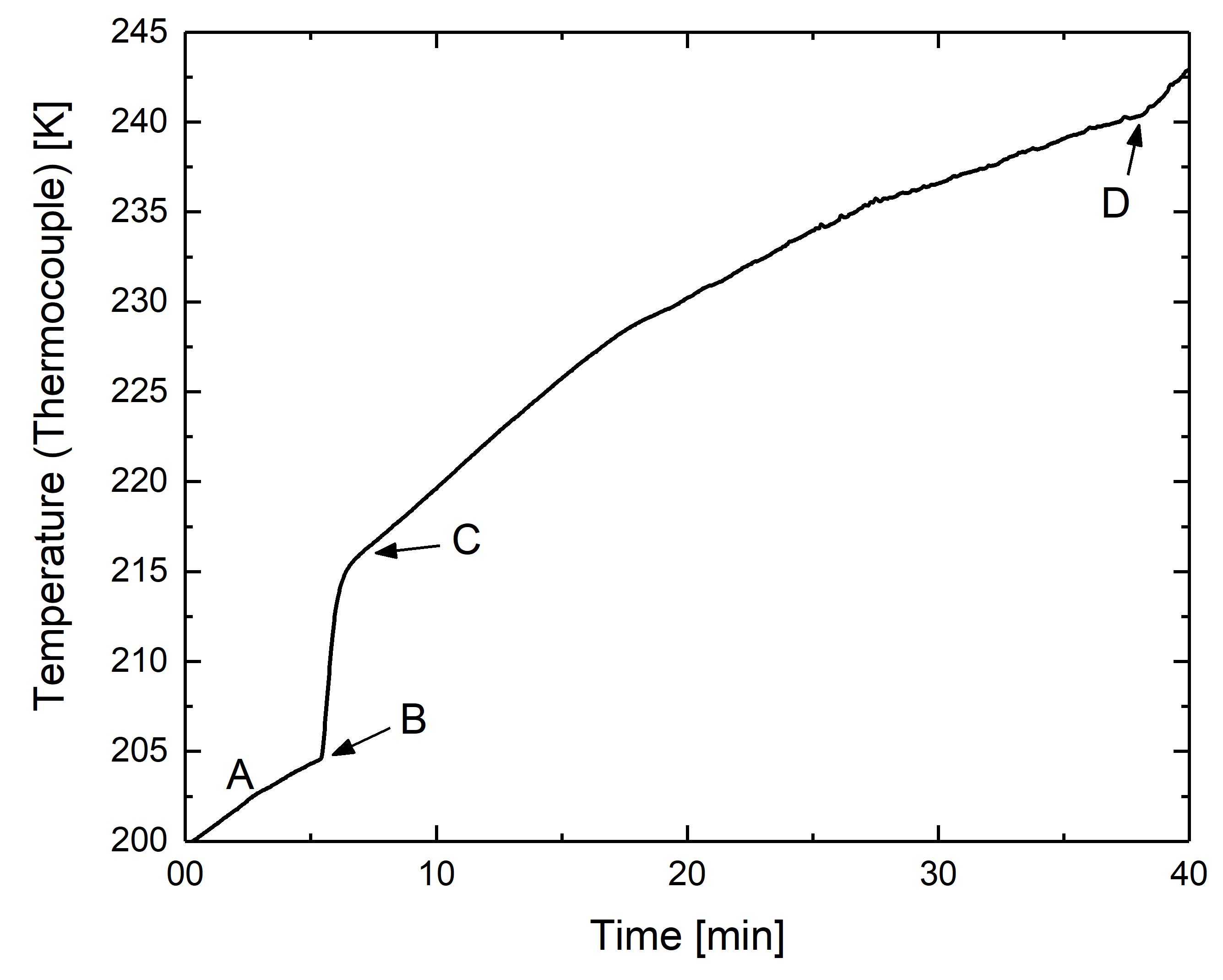}
    \caption{Temperature of the ice measured by the upper thermocouple sensor during an experimental run (dust aggregate size: $(0.173 \pm 0.048 )  \, \mathrm{mm}$ in diameter; thickness: $3.9 \pm 1.2$ monolayers; see also Figure \ref{fig_7}). Phase A describes the passive warm-up of the sample. The heater was switched on at point B and at point C the dust activity started. Point D marks the end of dust activity.}
    \label{fig_4}
\end{figure}

To avoid condensation of frost during sample transfer, the ice-aggregate sample was covered by a plastic lid. Before insertion into the vacuum chamber, the sample was cooled down to a temperature $\lesssim 180$~K using liquid nitrogen so that no spontaneous dust ejection was possible at the installation of the sample and at the start of the measurements. After the sample had reached its desired temperature, it was rapidly transferred into the experimental setup, the thermocouples and heater cables were connected, and evacuation of the chamber was started. The sample was not actively cooled and, thus, the ice temperature slowly increased with time. Using the heater, the temperature could be steadily increased to about $250\, \mathrm{K}$. Temperature sensors and cameras were used to continuously monitor the temperature and surface evolution during the experimental runs. Figure \ref{fig_4} presents a typical temperature evolution during an experimental run.
\begin{figure}
	\includegraphics[width=\columnwidth]{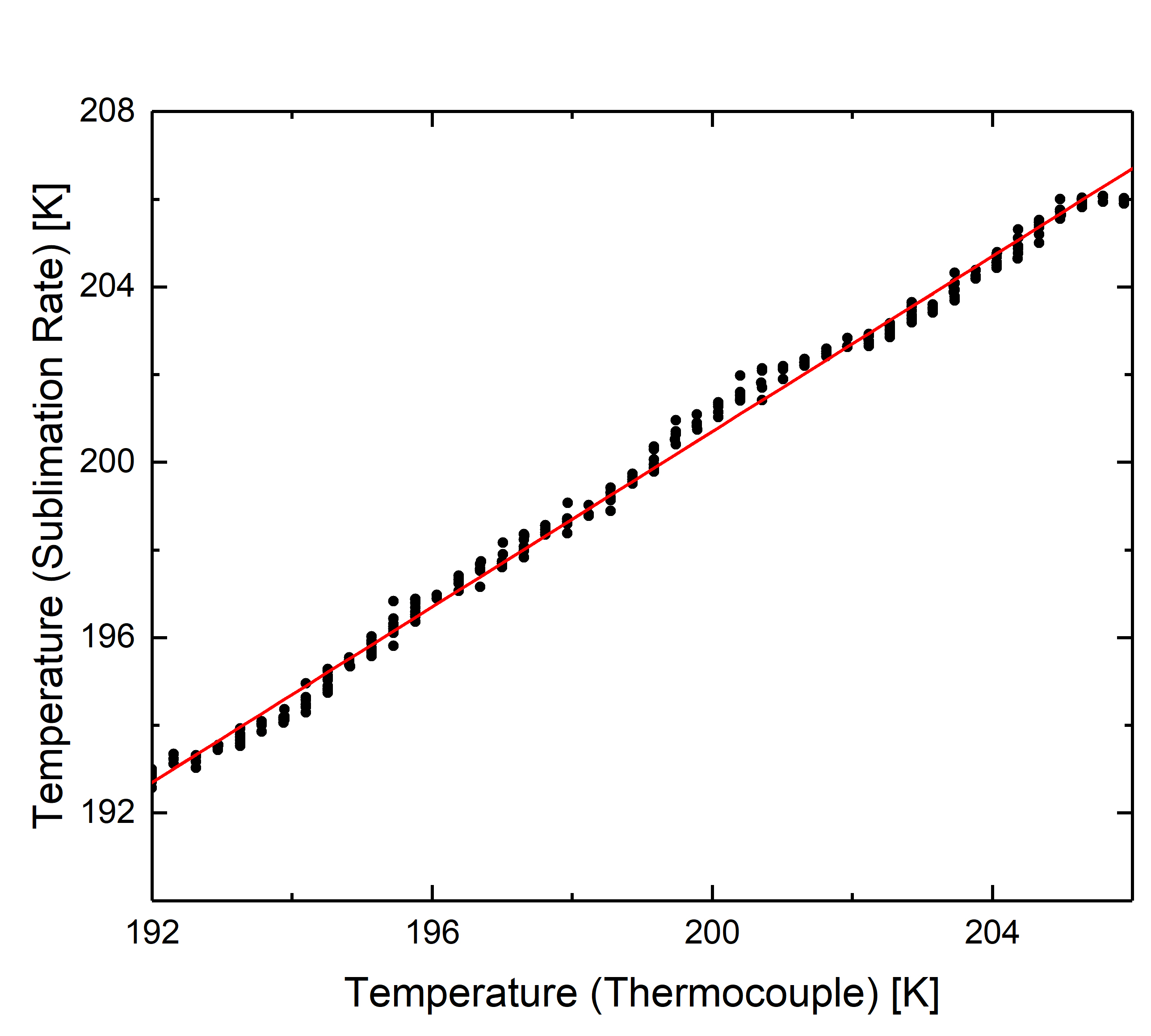}
    \caption{Water-ice temperature derived by the sub-surface thermocouple measurements (x-axis) and the surface temperature inferred using the outgassing rate of the sample (y-axis). The red curve shows the resulting fit function, shown in equation \ref{eq_temp}.}
    \label{fig_5}
\end{figure}
\begin{figure*}
	\includegraphics[width=\textwidth]{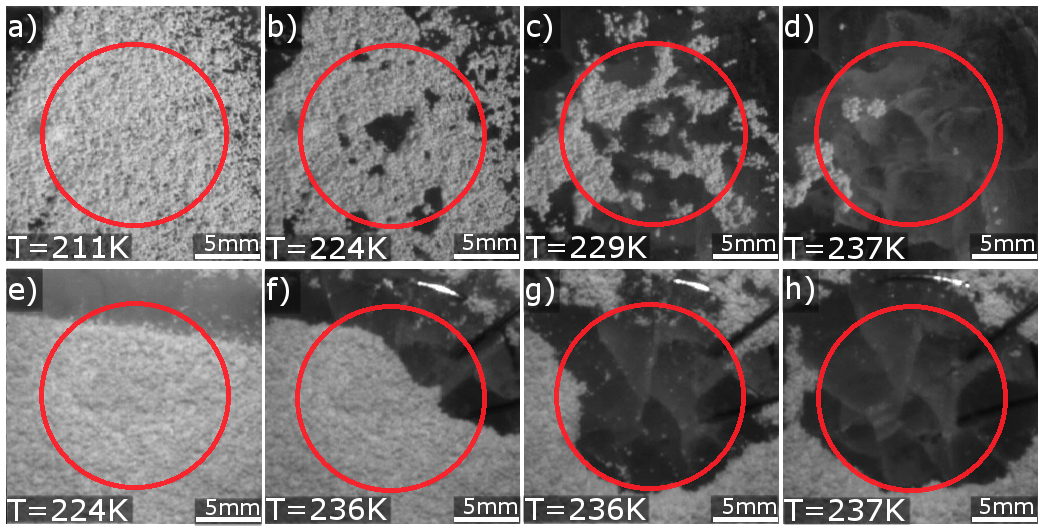}
    \caption{Image sequences of two samples covered by dust aggregates with $(0.173 \pm 0.048 )  \, \mathrm{mm}$ diameter. Top row (a-d): thickness $4.9 \pm 1.4$ monolayers. Bottom row (e-h): thickness $8.3 \pm 2.4$ monolayers. The red circle shows the area in which the fractional dust-cover data was acquired. Videos of the two experiments are available in the online version of this article.}
    \label{fig_6}
\end{figure*}

\subsection{Temperature Calibration}
During the experiment campaign, the question arose whether the upper temperature sensors measured the exact surface temperature. The thermocouple sensor can be positioned close, but not exactly at the surface of the sublimating ice. However, there are two possibilities to solve this problem. First, the temperature can be measured at two different depths. Assuming a linear temperature profile inside the ice, which is a good assumption for a solid material, the surface temperature can be derived. However, this method does not take the boundary condition, i.e., sublimation cooling and radiative heat exchange of the surface with warm surroundings into account. Thus, we used another experimental setup, which was designed to investigate the thermophyiscal properties of sublimating dust-ice mixtures. This experimental setup is only roughly outlined here and will be described in great detail in a forthcoming publication. The basic principle of this experiment is to measure the sublimation rate of ice samples at low pressures and low temperatures by a mass spectrometer. The knowledge of the outgassing rate directly provides the surface temperature via the so-called Hertz-Knudsen-formula \citep{Knudsen1909}. For this calibration experiment, we used the same sample holder and we produced the sample exactly by the same procedure as for the nominal experiments. The temperature obtained from the upper thermocouple in the ice can then be correlated with the temperature derived by the sublimation rate, which yields the surface temperature of the ice sample. Figure \ref{fig_5} shows the correlation between the temperature measured by the upper thermocouple $T_\mathrm{sensor}$ and that inferred by sublimation $T_\mathrm{subl}$. These temperatures can be related by the simple equation
\begin{equation}\label{eq_temp}
T_\mathrm{subl}=T_\mathrm{sensor}+0.7\, \mathrm{K}\mathrm{,}
\end{equation}
which is shown in Figure 6 as a solid red line.

\section{Results}\label{Sect_3}
Two different perspectives were chosen to observe the gas-driven dust activity with the cameras. A view from top enabled the observation of the surface evolution of the samples. Additionally, the samples were observed from the side, which allowed the observation of the aggregate trajectories during ejection.

\subsection{Surface Evolution}

\begin{figure}
	\includegraphics[width=\columnwidth]{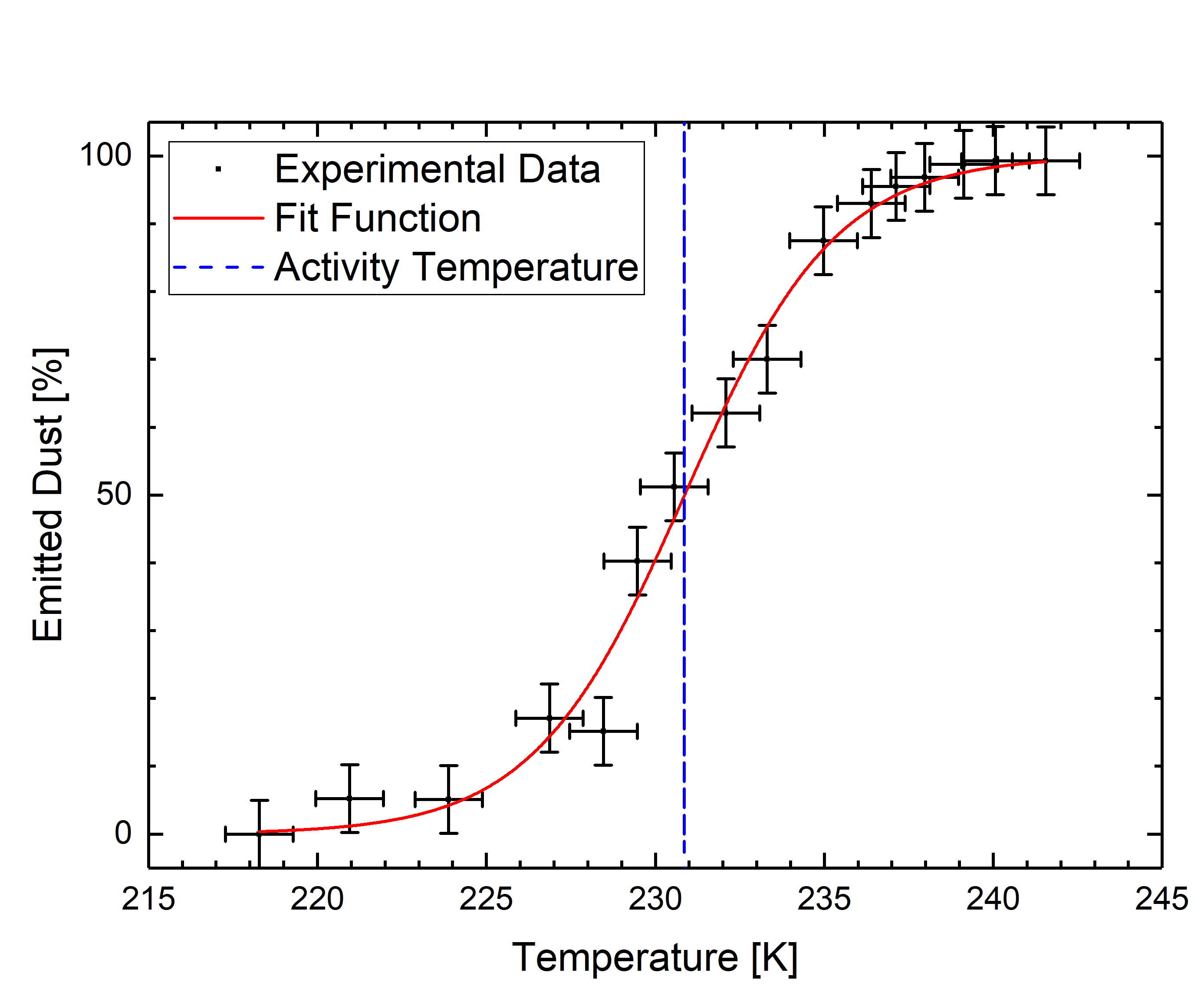}
    \caption{Example for the temperature dependence of the fractional dust cover for dust aggregates with $(0.173 \pm 0.048 )  \, \mathrm{mm}$ diameter and a thickness of $3.9 \pm 1.2$ monolayers.  For comparison, the fit function (Eq. \ref{eq:boltzmann}) and the resulting activity temperature are shown by the red curve and the blue dashed line, respectively.}
    \label{fig_7}
\end{figure}

When the surface of the ice had reached a certain temperature, the dust layer visibly began to erode, due to the ejection of dust aggregates. The dust cover appears white in the camera view from the top, whereas the transparent water ice is dark. Images were taken in appropriately chosen time steps to ensure a good data resolution with respect to the temperature of the sample. Figure \ref{fig_6} shows two image sequences with the same dust-aggregate sizes ($(0.173 \pm 0.048)  \, \mathrm{mm}$ in diameter), but for different dust layer thicknesses ($4.9 \pm 1.4$ and $8.3 \pm 2.4$ monolayers, respectively). The fractions of the dust-covered area (i.e., white pixels in the binarized camera images) and of the dust-free area (i.e., black pixels in the binarized camera images) were derived in the central part of the sample (indicated by the red circles in Figure \ref{fig_6}) as a function of temperature of the water-ice surface. Figure \ref{fig_7} shows an example for the temperature dependence of the fractional dust cover of a sample with dust aggregates of $(0.173 \pm 0.048)  \, \mathrm{mm}$ in diameter and with a thickness of $3.9 \pm 1.2$ monolayers. The resulting fractional dust cover was fitted by a Boltzmann-function (solid red curve in Figure \ref{fig_7})
\begin{equation}
Q(T)=100-\frac{100}{1+\exp\left(\frac{T-T_{A}}{\Delta T}\right)}\mathrm{.}
\label{eq:boltzmann}
\end{equation}
We define the activity temperature $T_{A}$ as the temperature at which $50 \, \%$ of the dust layer within the red circle had been ejected. The parameter $\Delta T$ describes the sharpness of the transition from dust-covered to uncovered phases. Table \ref{Table:TADT} lists the derived values of $T_{A}$ and $\Delta T$.

\par
\begin{figure}
\includegraphics[angle=0,width=1.\columnwidth]{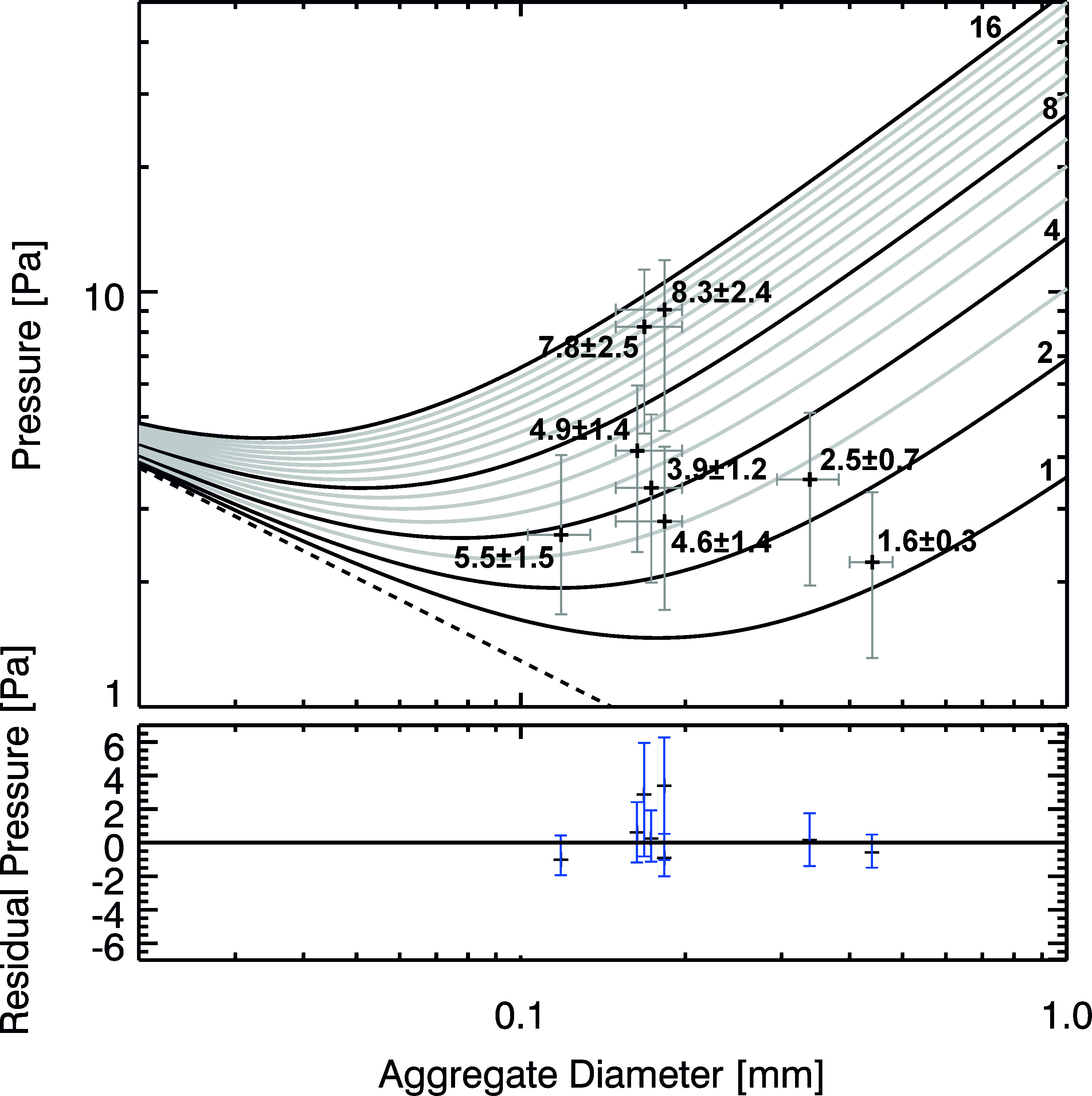}
    \caption{Derived gas pressures at the ice-dust interface when the fractional dust cover was 50\% (data points with error bars). For comparison, the pressures required to lift layers of dust aggregates are shown by the solid curves. The numbers next to the data points and to the solid curves denote the thickness of the dust-aggregate layers above the water-ice surface in units of aggregate diameters. For clarity, the five data points corresponding to dust-aggregate diameters of 0.173~mm are slightly offset in horizontal direction to avoid too much overlap. The uncertainty in pressure was estimated by the sharpness of temperature $\Delta T$ (see Equation \ref{eq:boltzmann} and Figure \ref{fig_7}). The lower panel shows the pressure residuals derived by subtracting the theoretical prediction (the curve derived for the respective number of layers as denoted by the numbers next to the data points) from the measured pressure.}
    \label{fig_8}
\end{figure}

From the measured activity temperature and from the estimated layer thickness, we derived the water-vapor pressure below the dust cover, i.e. $p_s ~ (1 - \xi)$ (see Equation \ref{Eqation_1} in Section \ref{Sect_1}). Figure \ref{fig_8} shows the derived pressures at the ice-dust interface as a function of dust-aggregate diameter used in the experiments (labeled data points with error bars). The dust-aggregate layer thickness in units of aggregate diameters is denoted by the numbers next to the data points. As shown by Equation \ref{Eqation_1}, the water-vapor pressure has to exceed the tensile strength of the material plus the gravitational load of the dust layers to be able to eject the dust aggregates (see curves in Figure \ref{fig_8}; the corresponding aggregate layers assumed for the calculations are shown by the numbers next to the curves). The residual pressure shown in the lower panel is derived by subtracting the theoretical prediction (the curve derived for the respective number of layers as denoted by the number next to the data point) from the measured pressure. It can clearly be seen that there are no systematic deviations between prediction and measurement.

\begin{table}
\centering
\caption{Sample types and resulting activity values according to Equation \ref{eq:boltzmann}, where $T_{A}$ and $\Delta T$ denote the activity temperature and sharpness, respectively.}
\label{Table:TADT}
\begin{tabular}{|c|c|c|c|}
\hline
Aggregate Diameter  & Layer Thickness & $T_{A}$  & $\Delta T$ \\
{[$\mathrm{mm}$]} & [monolayer] & [K] & [K]\\
\hline
$0.119 \pm 0.031$                & $5.5 \pm 1.5$   & $226.3$ & $2.3$     \\
 \hline
$0.173 \pm 0.048$                & $3.9 \pm 1.2$   & $230.9$ & $2.2$     \\
\hline
$0.173 \pm 0.048$                & $4.6 \pm 1.4$   & $231.8$ & $1.6$     \\
\hline
$0.173 \pm 0.048$                & $4.9 \pm 1.4$   & $228.1$ & $2.1$     \\
\hline
$0.173 \pm 0.048$                & $7.8 \pm 2.5$   & $235.4$ & $0.9$     \\
\hline
$0.173 \pm 0.048$                & $8.3 \pm 2.4$   & $236.0$ & $0.1$     \\
\hline
$0.338 \pm 0.088$                & $2.5 \pm 0.7$   & $234.0$ & $1.9$     \\
\hline
$0.440 \pm 0.079$                & $1.6 \pm 0.3$   & $232.9$ & $1.7$    \\ \hline
\end{tabular}
\end{table}
For the derivation of the tensile strength, we used the model developed by \citet{Skorov2012} which is supported by the laboratory results obtained by \citet{Blum2014} and \citet{Brisset2016}. On Earth, the gravitational load of the dust-aggregate layers is slightly higher than the tensile strength of the material (see Figures \ref{fig_1} and \ref{fig_8}), so that our experiments were not performed in the cohesive regime, but in the transition zone between the gravitational and the cohesive case. However, the data show (by comparing the numbers of the data points with the corresponding numbers of the curves in Figure \ref{fig_8}) a good match to the predictions. This means that Equation \ref{Eqation_1} is a reasonable approximation for the ice-sublimation-driven dust activity. Errors are due to measurement errors in temperature, aggregate diameter and layer thickness. The aggregate diameter and layer thickness show also a natural variation. Errors in calibration and deviation due to processes in the ice, e.g. latent heat, result in uncertainties of the temperature values.
\par
\begin{figure}
	\includegraphics[width=\columnwidth]{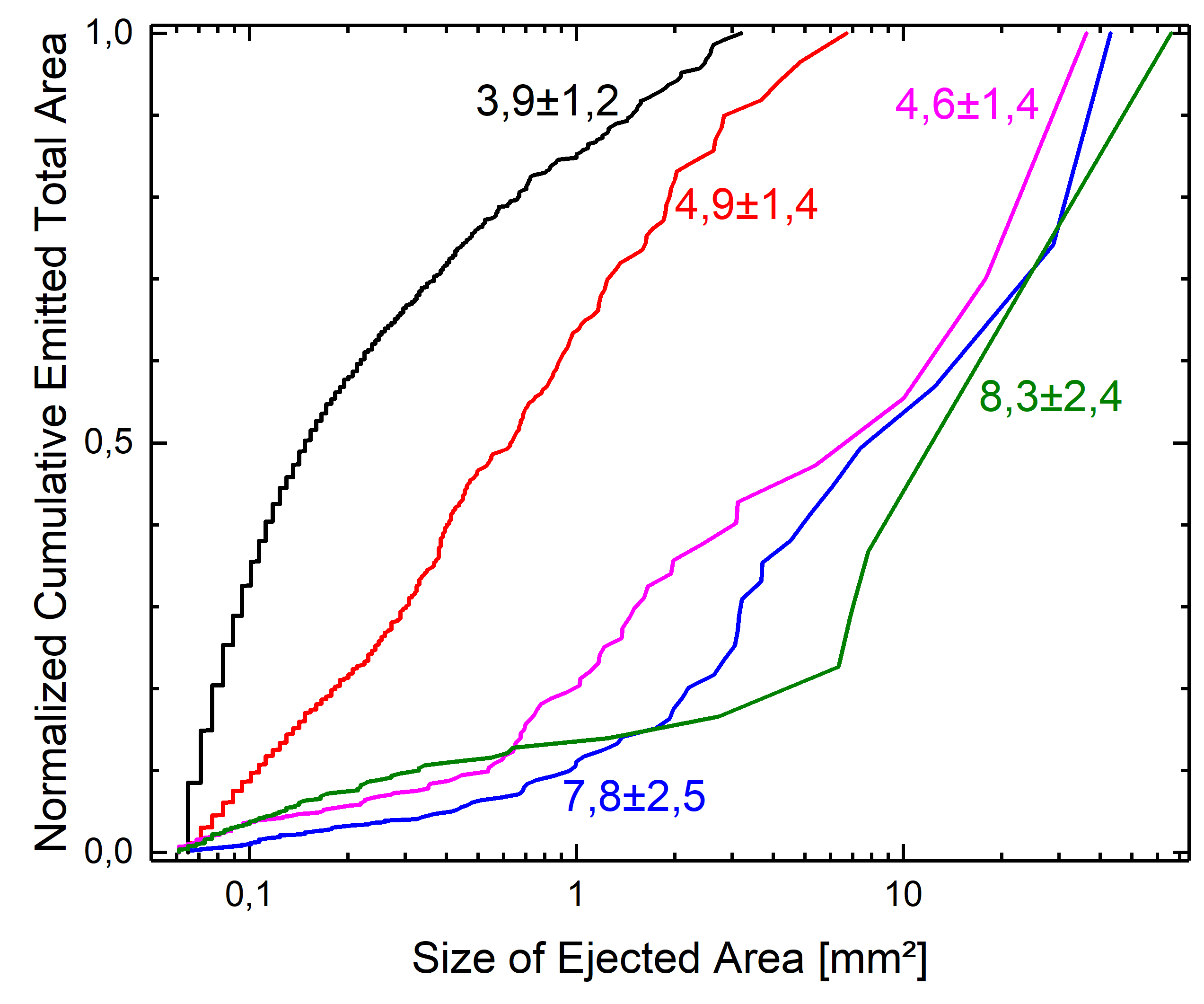}
    \caption{Normalized cumulative total area of emitted dust aggregates as a function of the size of the ejected clusters. The different curves show data from five different experiments with the same dust-aggregate diameter of $(0.173 \pm 0.048)  \, \mathrm{mm}$. The numbers next to the curves denote the thickness of the aggregate layer above the water-ice sample in units of aggregate diameters.}
    \label{fig_9}
\end{figure}
\begin{figure}
	\includegraphics[width=\columnwidth]{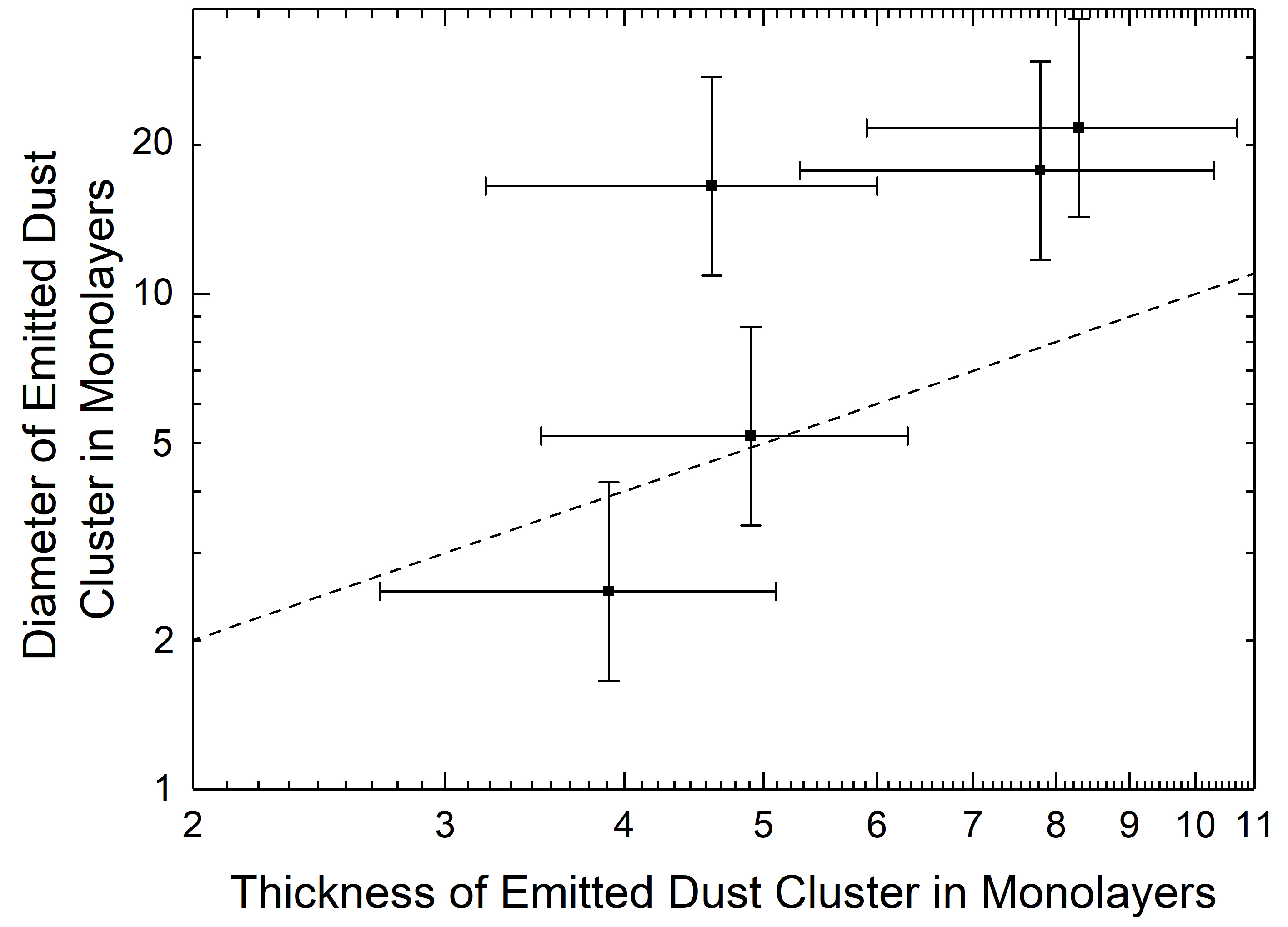}
    \caption{Diameter of the emitted dust cluster as a function of its thickness (both in units of the dust-aggregate diameter) for dust aggregates with diameters of $(0.173 \pm 0.048) \, \mathrm{mm}$. The vertical error bars correspond to one standard deviation from the median value of the total area of emitted dust aggregates, as shown in Figure \ref{fig_9}. The dashed curve is the expected cluster diameter if the clusters are assumed to be cylindric in shape with diameter equal to thickness.}
    \label{fig_10}
\end{figure}

Additionally, we observed that mostly clusters of dust aggregates were ejected by the active water-ice surface and not single dust aggregates. From the image analysis it was obvious that always the entire dust-aggregate layer was blown off by the sublimating water ice. Using the top view of the camera, we  determined the size of the ejected clusters for different layer thicknesses, but for a fixed dust-aggregate diameter of $(0.173 \pm 0.048)  \, \mathrm{mm}$. For this purpose, we compared the images of the sample surface in time steps of one second and determined the connected areas of those ejected clusters whose pixels became black in the next time step. From this analysis, we derived the normalized cumulative emitted total area as a function of the ejected cluster area and found a clear dependency on the aggregate-layer thickness (Figure \ref{fig_9}). From these distributions, we calculated the average emitted cluster diameter (assuming cylindrical cluster shape), which we plotted in Figure \ref{fig_10} as a function of cluster thickness. For comparison, we also plotted in Figure \ref{fig_10} as a dashed curve the expected cluster diameter if the clusters are assumed to be cylindric in shape with diameter equal to thickness. Although the measurement uncertainties are rather large, it seems that the diameter of the lifted dust clusters grows stronger than linear with increasing layer thickness. For instance, at a layer depth of 8 dust-aggregate diameters ($\sim 1.4 ~\mathrm{mm}$), the average cluster diameter is approximately 20 dust-aggregate diameters (see Figure \ref{fig_10}), which corresponds to a linear dimension of $\sim 3~ \mathrm{mm}$, which is roughly a factor of two larger than the depth.

\subsection{Aggregate trajectories}
\begin{figure}
	\includegraphics[width=\columnwidth]{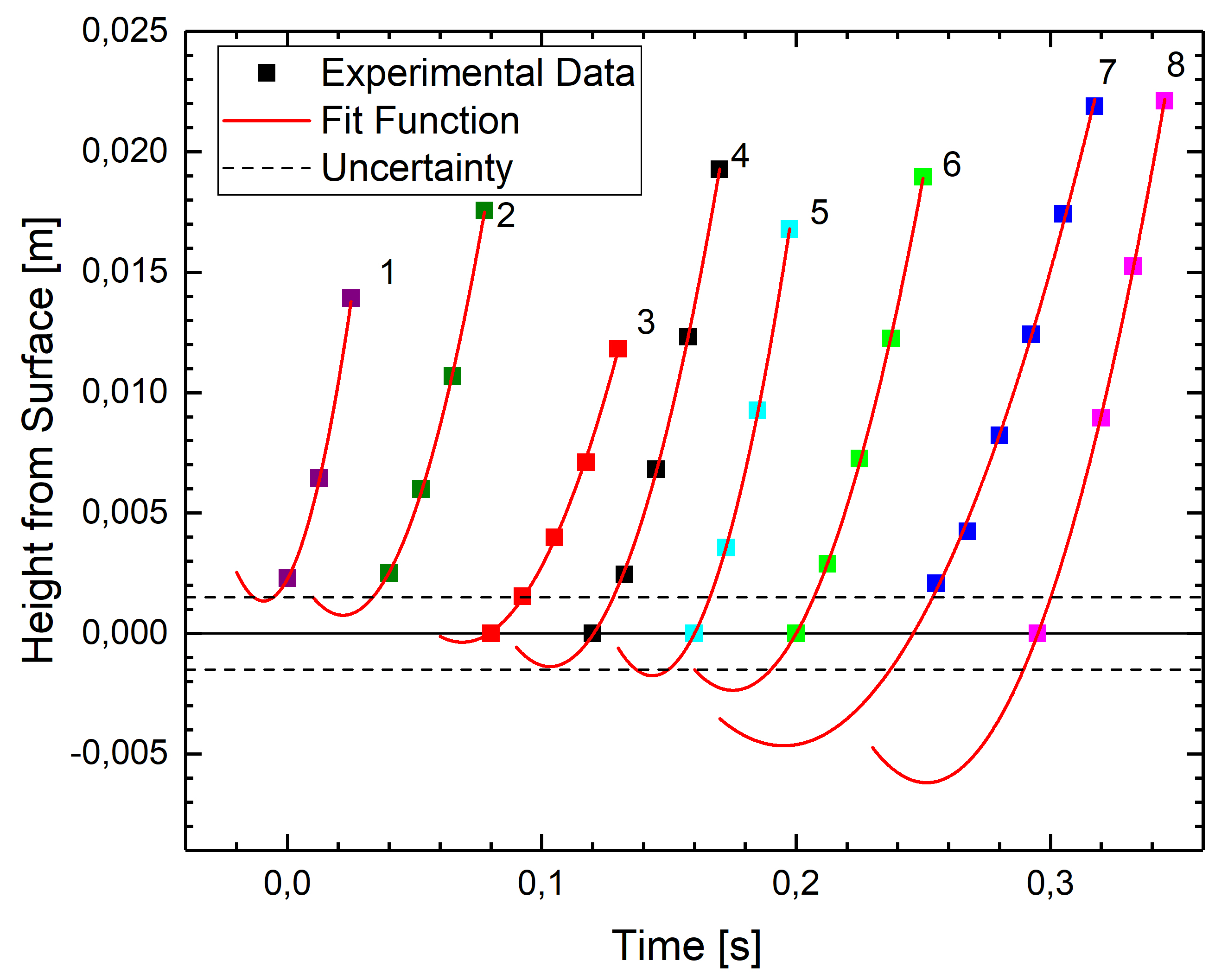}
    \caption{Observed trajectories of the emitted dust aggregates, or clusters and the fitted parabolic functions. The aggregates are labeled from left (1) to right (8). See Table \ref{initial_velocity} for details about the fit parameters. The dashed lines show the estimated uncertainty of the starting height.}
    \label{fig_11}
\end{figure}

\begin{table}
\centering
\caption{Fit parameters, including the initial velocity, obtained by fitting parabolic functions to the aggregate trajectories.Because the particle no. 1 was only measured three times, the parabolic function and thus the acceleration has no uncertainty.}
\label{initial_velocity}
\begin{tabular}{|c|c|c|}
\hline
 Particle  &  Initial  & Accelerations
\\
No. & Velocity {[m/s]} &  [m/$\mathrm{s}^2$]
\\ \hline
1 & $0^{+0,32}_{-0}$        & $10.55 $
\\ \hline
2 & $0^{+0,26}_{-0}$        & $5.41 \pm 0.69$
\\ \hline
3 & $0.07^{+0,08}_{-0,07}$  & $3.22 \pm 0.30$
\\ \hline
4 & $0.16^{+0,08}_{-0,07}$  & $4.64 \pm 0.25$
\\ \hline
5 & $0.21^{+0,08}_{-0,13}$  & $6.37 \pm 0.19$
\\ \hline
6 & $0.19^{+0.05}_{-0.08} $  & $3.80 \pm 0.30$
\\ \hline
7 & $0.22^{+0.06}_{-0.03}$  & $1.79 \pm 0.48$
\\ \hline
8 & $0.28^{+0.03}_{-0.04}$  & $3.23 \pm 0.34$
\\ \hline
\end{tabular}
\end{table}

The side view of the samples allowed us to observe the trajectories of single aggregates or clusters when ejected by the outflowing water molecules. In total, the trajectories of eight particles were recorded (see Figure \ref{fig_11}) and fitted by parabolic functions to derive the initial velocity and mean accelerations of the aggregates (see Table \ref{initial_velocity} for details). Assuming an uncertainty of our surface determination to be $1.5 \, \mathrm{mm}$, we can also derive upper and lower limits for the ejection speed. As indicated in Table \ref{initial_velocity}, two of the observed dust aggregates possessed no measurable initial velocity, but the other six aggregates had initial velocities exceeding zero.

\section{Conclusion and applications to comets}\label{Sect_4}
The experimental results provide major implications for our understanding of cometary activity.
\par
First, the ejection of dust aggregates by sublimating water ice as measured in the laboratory shows that cometary activity is possible exactly as predicted by \citet{Gundlach2015} under the assumption that comets formed by the gravitational collapse of an ensemble of dust aggregates and are, thus, composed of mm- to cm-sized dust pebbles \citep[see][for details]{Blum2014,Blum_2017}. Our experiments show that dust is emitted off a sublimating ice surface if the pressure build-up below the dust is stronger than the cohesive strength of the dust. The latter derives from the combination of adhesion and gravitational force. With this result, we confirm the dust emission model described in Equations \ref{Eqation_1}-\ref{Eqation_3} (see Figure \ref{fig_8}). To apply these results to real comets, Equation \ref{Eqation_2} needs to be adapted to the cometary surface acceleration, as shown in Figure \ref{fig_1}. As one can see in Figure \ref{fig_1}, pebbles of dm size are the easiest to detach, which is in full agreement with the observations of comet 67P analyzed by \citet{Blum_2017} (their Figure 7). Although our experiments used solid ice under a desiccated dust cover, our results are still applicable to comets, because as stated in the dust ejection model (Equations \ref{Eqation_1}-\ref{Eqation_3}), the responsible parameter is the water vapor pressure, which is only dependent on the water-ice temperature and not on the water-ice abundance. Thus, our results are fully applicable to cometary nuclei by extrapolating our findings to their low-gravity regime (see, e.g., Figure \ref{fig_1} for 67P). It should, however, be noted that the water-ice temperature on a real comet may be influenced by a variety of parameters, such as the water-ice abundance or the mixing mode of water ice and refractory materials.
\par
Second, the size of the ejected dust-aggregate clusters depends on the thickness of the dust cover. The thicker the dust layer, the larger the ejected dust-aggregate clusters are. This process can explain the large chunks observed in the inner coma of comets 103P/Hartley 2 \citep{AHearn2011} and 67P/Churyumov-Gerasimenko \citep{Thomas2015,Rotundi2015} together with the assumption that comets are made of mm- to cm-sized dust pebbles. With the relation between the size of the emitted dust clusters and the depth of the desiccated dust-aggregate layers, as shown in Figure \ref{fig_10}, it will be in principle feasible to determine the depth at which the dust-ice interface is located if a proper measurement of the average size and shape of the emitted dust can be performed. However, this might be a difficult task, because it is still unclear at which dominant size most of the dust mass is emitted. For instance, \citet{Blum_2017} showed that for comet 67P/Churyumov-Gerassimenko this size falls in the vast range of $\sim 1 ~\mathrm{mm}$ to $\sim 10 ~\mathrm{m}$ and \citet{Fulle2016b} and \citet{ott2017} derived mass-frequency distribution functions with peak masses of $\sim 1 \,\mathrm{kg}$ and, thus, typical dimensions of $10 \,\mathrm{cm}$. Future work might narrow this range down and allow a stricter assessment of the depth of the dust-ice boundary. As can be seen in Figure \ref{fig_10}, the horizontal dimensions exceed the vertical of the emitted clusters by roughly a factor of 2 so that the escaping chunks can most likely be described as oblate spheroids. When the dust-ice interface is at shallow depths as, e.g., assumed by \citet{Blum_2017}, the emitted dust chunks cannot be much lager than a few aggregates in diameter. Larger dust chunks can only be emitted if the sublimation front can be found at higher depths, e.g., if the activity is driven by CO$_2$. It should be mentioned that our studies concentrated on a well-defined boundary between water ice and silica aggregates and that other materials and mixtures of ice may influence the adhesion between the aggregates and thus the required temperature for dust ejection.
\par
Third, the observation of the aggregate trajectories provided evidence for an initial starting velocity greater than zero, which the aggregates, or clusters, must have obtained during the ejection event. The nature of the process leading to a non-vanishing initial velocity of the aggregates cannot be revealed in this work. However, \citet{Ratke1989} observed vibrating particles in the KOSI experiments and concluded that this might lead to the non-zero starting velocity. It is interesting to note that an initial velocity greater than zero had to be assumed by \citet{Kramer2015} to accurately model the observed inner coma structure of comet 67P.

\section{Summary and perspectives}
In this paper, we present our first comet-simulation experiments on the ejection of dust aggregates from a sublimating water-ice surface (see Section \ref{Sect_2}). We performed eight experiments, which were used to study the temperature at which ejection of dust aggregates by sublimation of water ice occurs (see Section \ref{Sect_3}). In these experiments, the surface evolution was measured by monitoring the fraction of dust and ice making up the sample surface. Additionally, the velocities of the ejecting aggregates were investigated.
\par
Based on the experiments we were able to draw three main conclusion about cometary activity (see Section \ref{Sect_4}):
\begin{enumerate}
\item Cometary dust aggregates can be emitted by the gas pressure build-up at the ice-dust interface. Dust emission only works for surfaces consisting of aggregates because of their very low tensile strengths. Homogeneous dust layers consisting of $\mathrm{\mu m}$-sized dust particles possess cohesive strengths orders of magnitude too high to cause dust activity.
\item The size of the ejected dust-aggregate clusters depends on the thickness of the dust cover. The thicker the dust layer, the larger the ejected dust-aggregate clusters are. This relationship allow the determination of the ice-dust interface depth if the size of the ejected chunks can be measured.
\item Ejected aggregates start with a non-zero initial velocity.
\end{enumerate}
The next step for the upcoming experiments is to use more realistic volatiles, such as granular water or carbon-dioxide ice. Furthermore, the installation of additional equipment will allow for a precise measurements of the outgassing rate and the composition of the gas. Additionally, experiments dedicated to study the origin of the  non-zero starting velocity of the aggregates will be performed.

\section*{Acknowledgements}
We aknowledge scientific contribution from the CoPhyLab project funded by the D-A-CH programme ((DFG GU 1620/3-1 and BL 298/26-1 / SNF 200021E\_177964 / FWF I 3730-N36).\\
This work was supported by the Deutsche Forschungsgemeinschaft (DFG) through grant GU 1620/1-1 and by the Deutsches Zentrum f\"ur Luft- und Raumfahrt (DLR) through grant 50WM1536.
\par
We thank Marc Pfannkuche for providing us with the images of the dust-aggregate samples and Yuri Skorov and Horst Uwe Keller for continuous fruitful discussions.

\bibliographystyle{mnras}
\bibliography{references}

\label{lastpage}
\end{document}